\documentclass[prb,twocolumn,amsmath,nofootinbib,amssymb,superscriptaddress]{revtex4}
\usepackage{graphicx}

\usepackage{epsfig,psfrag,amsmath,amssymb,float}
\begin{document}

\newcommand{\ltwid}{\mathrel{\raise.3ex\hbox{$<$\kern-.75em\lower1ex\hbox{$\sim$}}}}
\newcommand{\gtwid}{\mathrel{\raise.3ex\hbox{$>$\kern-.75em\lower1ex\hbox{$\sim$}}}}
\def\K{{\bf{K}}}
\def\Q{{\bf{Q}}}
\def\Gbar{\bar{G}}
\def\tk{\tilde{\bf{k}}}
\def\k{{\bf{k}}}
\def\kt{{\tilde{\bf{k}}}}
\def\p{{\bf{p}}}
\def\pp{{\bf{p}}^\prime}
\def\Gpp{\Gamma^{pp}} 
\def\Phid{\Phi_d(\k,\omega_n)} 
\def\ld{\lambda_d(T)}
\def\n{\langle n \rangle }
\def\dw{d_{x^2-y^2}}

\title{The Pairing Interaction in the 2D Hubbard Model}

\author{T.A.~Maier}
\affiliation{Computer Science and Mathematics Division,\\
 Oak Ridge National Laboratory, 
Oak Ridge, TN 37831-6164}
\email{maierta@ornl.gov}

\author{M.~Jarrell}
\affiliation{Department of Physics,\\
 University of Cincinnati, Cincinnati, OH 45221}
\email{jarrell@physics.uc.edu}

\author{D.J.~Scalapino}
\affiliation{Department of Physics,\\
 University of California, Santa Barbara, CA 93106-9530}
\email{djs@vulcan2.physics.ucsb.edu}

\date{\today}

\begin{abstract}

A dynamic cluster quantum Monte Carlo approximation is used to study the
effective pairing interaction of a 2D Hubbard model with a near neighbor hopping
$t$ and an on-site Coulomb interaction $U$ . The effective pairing interaction
is characterized in terms of the momentum and frequency dependence of the
eigenfunction of the leading eigenvalue of the irreducible particle-particle
vertex. The momentum dependence of this eigenfunction is found to vary as
$(\cos k_x-\cos k_y)$ over most of the Brillouin zone and its frequency
dependence is determined by the exchange energy $J$. This implies that the
effective pairing interaction is attractive for singlets formed between
near-neighbor sites and retarded on a time scale set by $J^{-1}$. The strength
of the pairing interaction measured by the size of the d-wave eigenvalue peaks
for $U$ of order the bandwidth $8t$. It is found to increase as the system is
underdoped.

\end{abstract}

\maketitle


\section{Introduction}

Results from numerical studies suggest that the two-dimensional (2D)
Hubbard model exhibits the basic phenomena which are seen in the
high-$T_c$ cuprate materials. At half-filling, one finds a groundstate
with long-range antiferromagnetic order \cite{Hir85}. Doped away from
half-filling there is a pseudogap regime
\cite{Mai01,Sta03,TKS05,Kyu04,Han05,Han05a}, and at low temperature a
striped phase \cite{Hag05, WS03} as well as d-wave pairing \cite{Mai05}.
In addition, the various phases appear delicately balanced with respect
to changes in parameters. All of these features remind one of the actual
cuprate materials, so that it is of interest to understand the structure
of the interaction that leads to $\dw$-wave pairing in this model.

Here we will study this interaction for a 2D Hubbard model \cite{And87}
on a square lattice which contains a one-electron near-neighbor hopping
$t$ and an onsite Coulomb interaction $U$. In this case the results
depend upon just two parameters $U/t$ and the average site occupation
$\n$. Previous work \cite{MJS06} has shown that the pairing interaction
with $U/t=4$ and $\n=0.85$ increases with momentum transfer, has a
Matsubara frequency dependence similar to that of the $\Q=(\pi,\pi)$
spin susceptibility and is mediated by a particle-hole $S=1$ exchange
channel. Here we extend this work to explore the pairing interaction for
larger values of $U/t$ and various dopings. We are particularly
interested in the case in which $U$ is of order of the bandwidth $8t$.
In this case, well developed upper and lower Hubbard bands are seen in
the single particle density of states
\begin{eqnarray}\label{eqn:DOS}
	N(\omega)=-\frac{1}{\pi N}\sum_\k {\rm Im}\,
	G(\k,i\omega_m\rightarrow \omega+i\delta)\, .
\end{eqnarray}
\begin{figure}[htpb]
	\begin{center}
 		\includegraphics[width=3.5in]{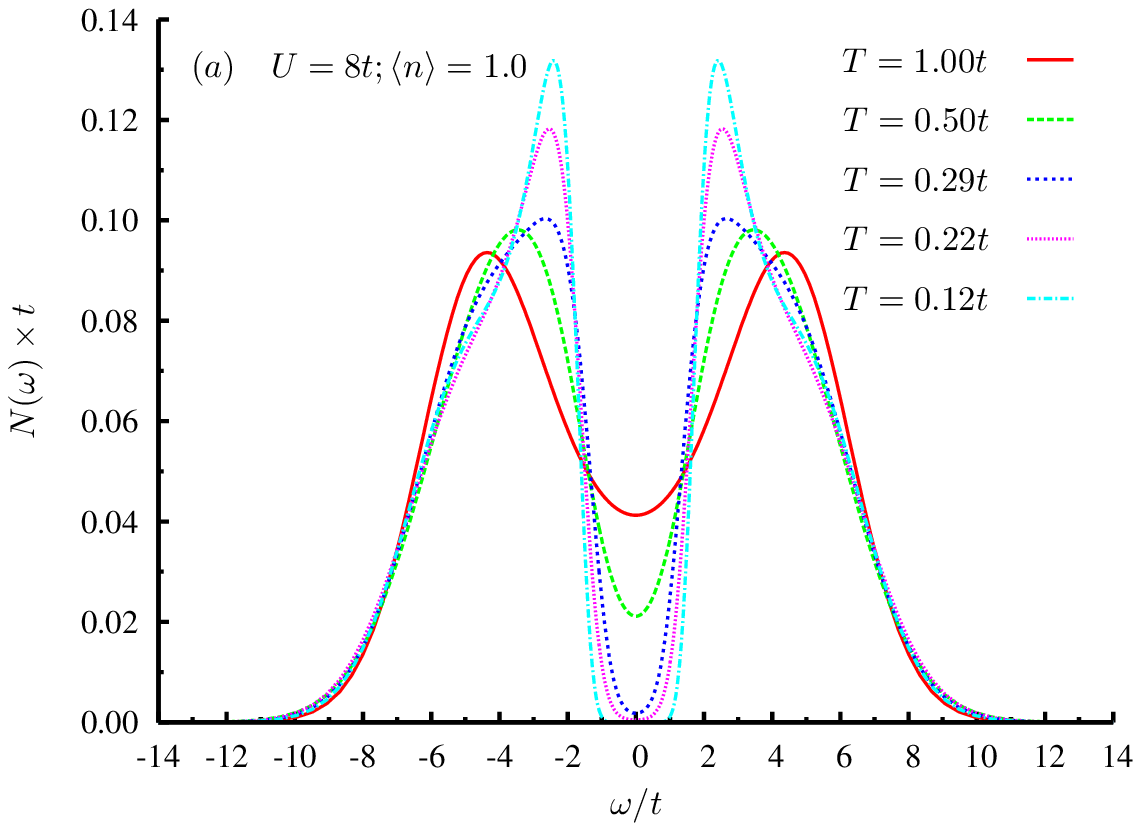}
 		\includegraphics[width=3.5in]{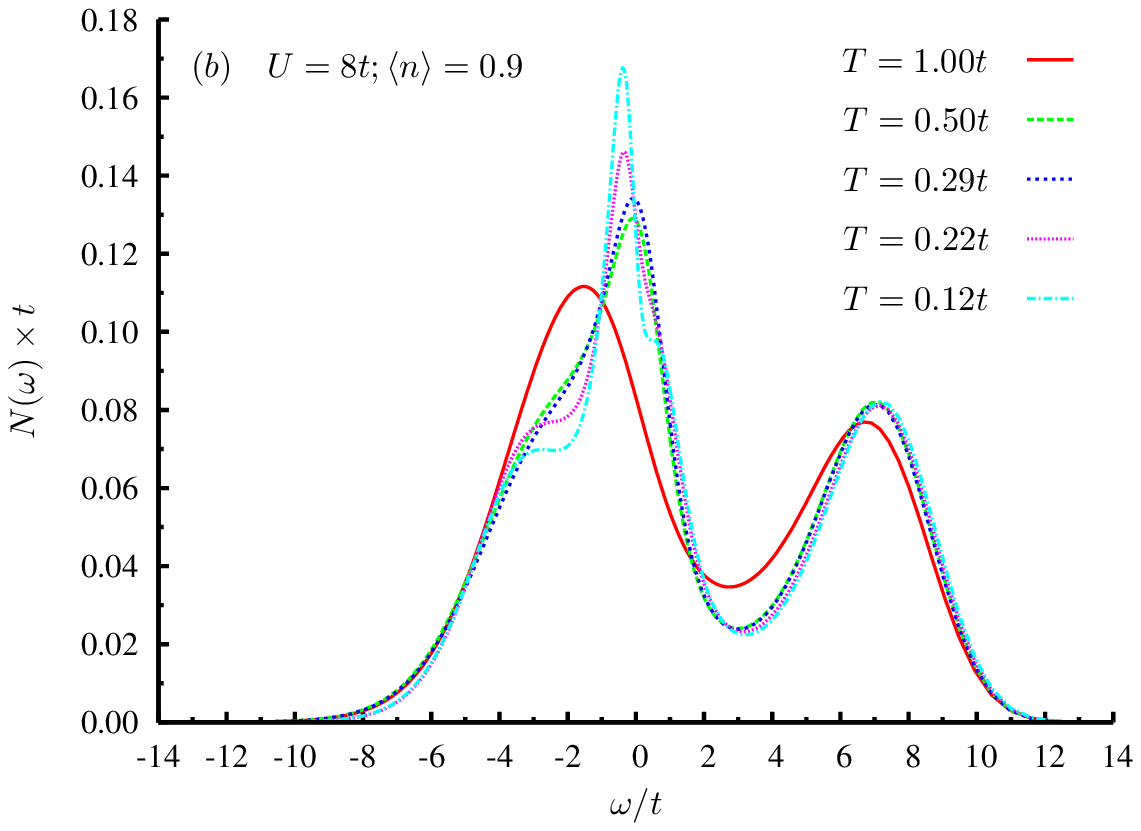}
	\end{center}
	\caption{The single particle density of states $N(\omega)$
	versus $\omega$ for $U=8t$, $N_c=15$ and various values of the
	temperature $T$ for site fillings of a) $\n=1.0$, b)
	$\n=0.90$.}
	\label{fig:DOS}
\end{figure}
Fig.~\ref{fig:DOS} shows $N(\omega)$ with $U/t=8$ for various
fillings. These results were obtained from a maximum entropy
continuation\cite{MEM} of dynamic cluster quantum Monte Carlo data on a 16 site
cluster. At half-filling, Fig.~\ref{fig:DOS}a, for $T\sim t$ one sees
broad upper and lower Hubbard bands. Then, as the temperature scale
drops below the exchange energy scale $J\sim 4t^2/U$, two additional
structures appear above and below $\omega=0$. As Preuss {\it et.~al}
\cite{Pre95} first showed, the inner structures arise from the formation of
two coherent bands, each of width $\sim 2J$ that form as the antiferromagnetic
correlations develop. This characteristic Mott-Hubbard-antiferromagnetic four
band structure \cite{Pre95,Mor95} is clearly seen when $U$ becomes of order or
larger than the bandwidth.

Fig.~\ref{fig:DOS}b shows $N(\omega)$ for the doped $\n=0.9$
case. Here the chemical potential has moved down into the lower
coherent band and the spectral weight in the upper coherent band has
essentially vanished. Nevertheless, the remnants of the upper and lower Hubbard
bands remain. Part of the motivation for this study is to examine the
structure of the pairing interaction in this parameter regime. 

\begin{figure}[htbp]
	\begin{center}
		\includegraphics[width=3.5in]{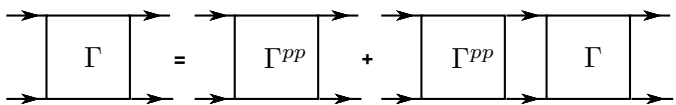}
	\end{center}
	\caption{The Bethe-Salpeter equation for the particle-particle
	channel showing the relationship between the four-point vertex
	$\Gamma$ and the particle-particle irreducible vertex $\Gpp$.
	The solid lines are dressed single particle Green's functions.}
	\label{fig:BSE}
\end{figure}

As in our previous work, the pairing interaction $\Gamma^{pp}$ will be
calculated using a dynamic cluster quantum Monte Carlo simulation \cite{Mai05a}. As
shown in Fig.~\ref{fig:BSE}, $\Gamma^{pp}$ is the irreducible part of
the 4-point particle-particle vertex in the zero center of mass and
energy channel. Previously, we discussed how one could extract
$\Gamma^{pp}(k|k')=\Gamma^{pp}(k,-k;k',-k')$ with
$k=(\k,i\omega_n)$ using a dynamic cluster quantum Monte Carlo
simulation. 
\vskip .10in

While we studied $\Gpp$ in our earlier work, here we will focus on the
momentum and Matsubara frequency dependence of the $\dw$-wave
eigenfunction $\Phid$ of the homogeneous particle-particle
Bethe-Salpeter equation
\begin{equation}
-\frac{T}{N}\ \sum_{k^\prime} \Gamma^{pp} (k|k^\prime)\, G_\uparrow (k^\prime)\,
G_\downarrow (-k^\prime)\, \Phi_\alpha (k^\prime) = \lambda_\alpha\Phi_\alpha
(k)\, .
\label{two}
\end{equation}
At low temperatures, $\Phid$ has the largest eigenvalue and this
eigenvalue goes to one at $T=T_c$. As $T$ approaches $T_c$, the
momentum and frequency dependence of $\Phid$ reflect the structure of
the pairing interaction at $T_c$, just as the superconducting gap
function reflects the $\k$ and $\omega$ dependence of the pairing
interaction in the superconducting state. Thus, while $\Phid$ is not a
quantity that is directly measurable, it has a $\k$-dependence related
to the momentum dependence of the interaction and a Matsubara
frequency dependence which decays beyond a characteristic frequency
associated with the dynamic character of the interaction. It also has
the great advantage of depending upon one momentum and frequency
variable as opposed to the multiple momentum and frequency variables
of $\Gpp(k|k')$.

In the following section~\ref{sec:dynam-clust-appr} we review the
dynamic cluster approximation and discuss how one calculates the
$\dw$-wave eigenvalue $\ld$ and eigenfunction $\Phid$. Then, in
Sec.~\ref{sec:struct-pair-inter} we investigate how $\ld$ depends upon
$U$ and $\n$. Following this, we examine the $\k$-dependence of
$\Phid$ and see how closely it follows the simple
$\cos(k_x)-\cos(k_y)$ dependence. If it were of this form over the
entire Brillouin zone, then it would imply a strictly near-neighbor
pairing interaction. Then we turn to the $\omega_n$-dependence which
reflects the dynamics of the pairing interaction and study its
dependence on $U$. In Sec.~\ref{sec:separ-repr-pair}, based upon the
results for $\Phid$, we construct a simple separable representation of
$\Gpp(k|k')$ and discuss the strength of the pairing
interaction. Sec.~\ref{sec:conclusion} contains our conclusions.

\section{The Dynamical Cluster Approximation}
\label{sec:dynam-clust-appr}

The Dynamical Cluster Approximation (DCA) \cite{Mai05a} maps the bulk
lattice to a finite size cluster embedded in a self-consistent bath
designed to represent the remaining degrees of freedom. Short-range
correlations within the cluster are treated explicitly, while the
longer-ranged physics is described by a mean-field. By increasing the
cluster size, the DCA systematically interpolates between the
single-site dynamical mean-field result and the exact result, while
remaining an approximation to the thermodynamic limit for finite
cluster size.

\begin{figure}[htpb]
	\begin{center}
          \includegraphics[width=1.75in]{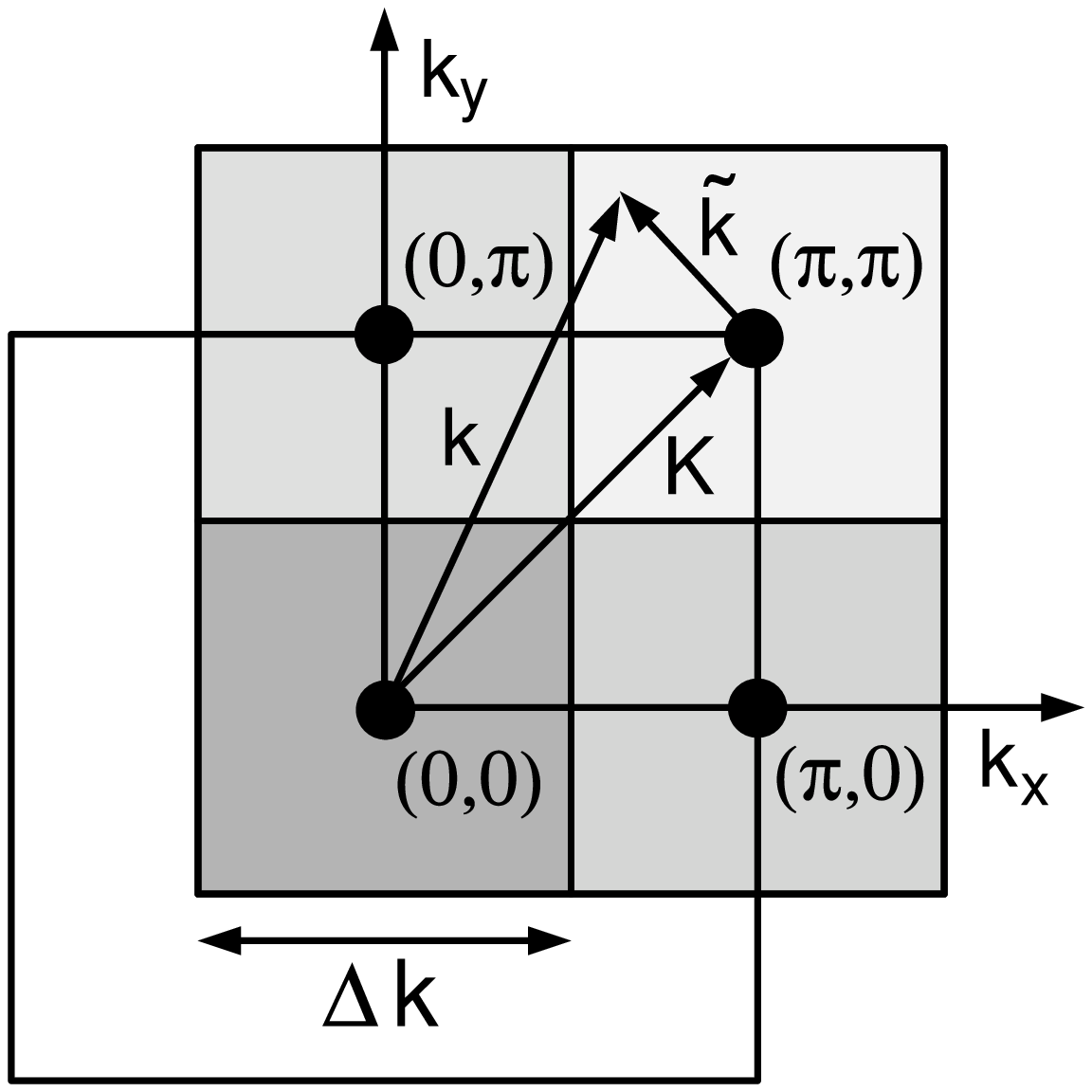}\includegraphics[width=1.75in]{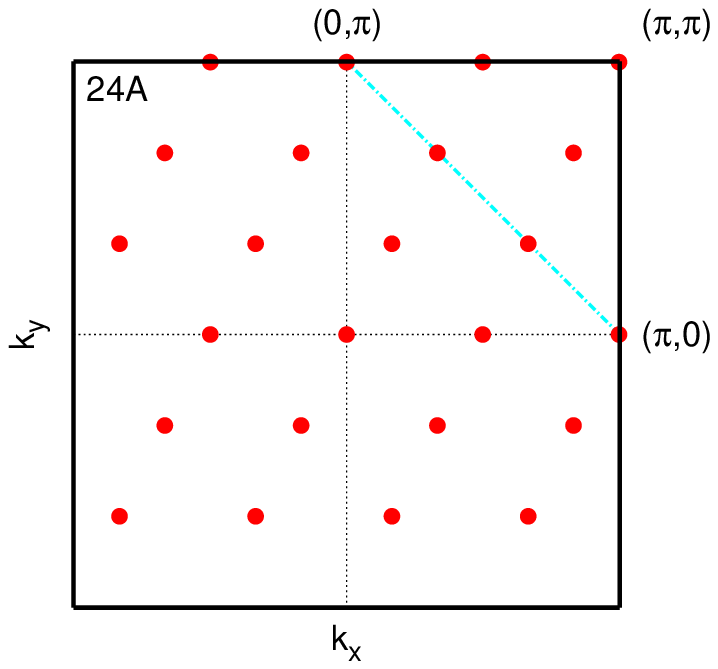}
	\end{center}
	\caption{In the DCA the Brillouin zone is divided into $N_c$
          cells each represented by a cluster momentum
          $\K$. Irreducible quantities such as the single-particle
          self-energy $\Sigma$ and two-particle irreducible vertex
          $\Gamma$ are constructed from coarse-grained propagators
          $\Gbar(\K)$ that are averaged over the momenta $\k'$ within
          the cell represented by $\K$. The cluster momenta $\K$ for a 4-site
	  cluster are shown on the left and those for a 24-site cluster on the right.}
	\label{fig:DCABZ}
\end{figure}

The essential assumption is that short-range quantities, such as the
single-particle self-energy $\Sigma$, and its functional derivatives,
the two-particle irreducible vertex functions, are well represented as
diagrams constructed from a coarse-grained propagator $\Gbar$. To
define $\Gbar$, the Brillouin zone in two dimensions is divided into
$N_c=L^2$ cells of size $2\pi/L^2$. As illustrated in
Fig.~\ref{fig:DCABZ}, each cell is represented by the cluster momentum
$\K$ in its center. The coarse-grained Green function $\Gbar(\K)$ is
then obtained from an average over the $N/N_c$ wave-vectors $\kt$
within the cell surrounding $\K$,
\begin{eqnarray}
  \label{eq:Gbar}
  \Gbar(\K,\omega_n) = \frac{N_c}{N}\sum_{\kt}
  \frac{1}{i\omega_n-\epsilon_{\K+\kt}
    +\mu-\Sigma_c(\K,\omega_n)}.
\end{eqnarray}
Here the self-energy for the bulk lattice $\Sigma(\K+\kt,\omega_n)$
has been approximated by the cluster self-energy
$\Sigma_c(\K,\omega_n)$.
Consequently, the compact Feynman diagrams constructed from
$\Gbar(\K,\omega_n)$ collapse onto those of an effective cluster
problem embedded in a host which accounts for the fluctuations arising
from the hybridization between the cluster and the rest of the system.
The non-interacting part of the effective cluster action is then
defined by the cluster-excluded inverse Green's function
\begin{eqnarray}
  \label{eq:Gscript}
  {\cal G}^{-1}(\K,\omega_n) = \Gbar^{-1}(\K,\omega_n)+\Sigma_c(\K,\omega_n)
\end{eqnarray}
which accounts for the hybridization between the cluster and the host. Given
${\cal G}^{-1}(\K,\omega_n)$ and the interaction on the cluster $U\sum_i
n_{i\uparrow}n_{i\downarrow}$, one can then set up a Hirsch-Fye quantum Monte
Carlo algorithm \cite{Hir86} to calculate the cluster Green function and from
it the cluster self-energy $\Sigma_c(\K,\omega_n)$ which is used in
Eq.~(\ref{eq:Gbar}) to re-calculate the coarse-grained Green function
$\Gbar(\K,\omega_n)$ \cite{Jar01,Mai05a}. This process is then iterated to
convergence.

Since a determinantal Monte Carlo method is used, there is also a sign
problem for the doped Hubbard model. However, the coupling of the
cluster to the self-consistent host significantly reduce the sign
problem so that lower temperatures can be reached \cite{Jar01}.

The DCA cluster one- and two-particle Green's functions that we
calculate have the standard finite temperature definitions
\begin{subequations}
\label{five}
\begin{equation}
  G_{c\sigma} (X_2, X_1) = 
  - \left\langle T_\tau c_\sigma (X_2)\, c^\dagger_\sigma (X_1)\right\rangle
\label{five a}
\end{equation}
and
\begin{eqnarray}
  G_{c 2\sigma_4\cdots\sigma_1} (X_4, X_3; X_2, X_1) = \hspace*{-4cm}&&\nonumber\\
  &-& \left\langle T_\tau c_{\sigma_4} (X_4)
    \, c_{\sigma_3} (X_3)\, c^\dagger_{\sigma_2} (X_2)\, c^\dagger_{\sigma_1} (X_1)\right\rangle
  \, .
\label {five b}
\end{eqnarray}
\end{subequations}
Here, $X_\ell = ({\bf X}_{\ell}, \tau_\ell)$, where ${\bf X}_{\ell}$
denotes a site in the DCA cluster, $\tau_{\ell}$ the imaginary time,
$T_\tau$ is the usual $\tau$-ordering operator, and
$c^{\dagger}_\sigma (X_2)$ creates a particle on the
cluster with spin $\sigma$.  Fourier transforming on both the cluster
space and imaginary time variables gives $G_c (K)$ and $G_{c2}(K_4,
K_3; K_2, K_1)$ with $K=({\K}, i\omega_n, \sigma)$.  Using $G_c(K)$
and $G_{c2}(K_4, K_3; K_2, K_1)$, one can extract the cluster
four-point vertex $\Gamma$ from
\begin{eqnarray}
\label{six}
G_{c2} (K_4, K_3; K_2, K_1) = \hspace*{-3cm} &&\nonumber\\
&-& G_c(K_1)\, G_c(K_2)\, \left[\delta_{K_1, K_4} \delta_{K_2, K_3}
  - \delta_{K_1, K_3} \delta_{K_2, K_4}\right]\nonumber \\
&+& \frac{T}{N}\ \delta_{K_1+K_2, K_3+K_4} G_c(K_4)\, G_c(K_3) \Gamma\, (K_4, K_3; K_2, K_1)\nonumber \\
&\times&  G_c(K_2)\, G_c(K_1)\, .
\end{eqnarray}
Then, using $G_c$ and $\Gamma$, one can determine the irreducible
particle-particle vertex $\Gamma^{pp}$ from the Bethe-Salpeter
equation shown in Fig.~\ref{fig:BSE}.

Using $\Gpp$, the eigenvalues and eigenfunctions of the Bethe-Salpeter
equation may then be calculated from
\begin{eqnarray}
  -\frac{T}{N}\ \sum_{k^\prime} \Gamma^{\rm pp} \left(K, -K; K^\prime, -K^\prime\right)\hspace*{-4cm}&&\nonumber\\
  &\times& G_\uparrow (k^\prime)\, G_\downarrow (-k^\prime)\, 
  \phi_\alpha (K^\prime) =
  \lambda_\alpha \phi_\alpha (K)
\label{seven}
\end{eqnarray}
Here, the sum over $k^\prime$ denotes a sum over both momentum
$\k^\prime$ and Matsubara $\omega_{n^\prime}$
variables. We decompose $\k^\prime = \K^\prime + \tk^\prime $.  By
assumption, irreducible quantities like $\Gamma^{pp}$ and
$\phi_\alpha$ do not depend on $\tk^\prime$, allowing us to
coarse-grain the Green function legs, yielding an equation that
depends only on coarse-grained and cluster quantities
\begin{eqnarray}
  -\frac{T}{N_c}\ \sum_{K^\prime} \Gamma^{\rm pp} \left(K, -K; K^\prime, 
-K^\prime\right)\,
  {{\bar{\chi}}_0^{\rm pp}}(K')
  \, \phi_\alpha (K^\prime) = &&\nonumber\\
  && \hspace{-2cm}\lambda_\alpha \phi_\alpha (K)
\label{Eq:eigvalcg}
\end{eqnarray}
with ${{\bar{\chi}}_0^{\rm pp}}(K') = \frac{N_c}{N}
\sum_{\tk^{\prime}} G_\uparrow (\K^\prime+\tk^\prime,i\omega_{n'}) \, G_\downarrow
(-\K^\prime-\tk^\prime,-i\omega_{n'})$.

Here, we show a number of
results for the 4-site cluster shown on the left in
Fig.~\ref{fig:DCABZ}, which allows us to investigate larger values of
$U$ and lower temperatures than the 24-site cluster. As discussed in
Ref.~\cite{Mai05}, the 4-site cluster does not allow for the effect of
pairfield phase-fluctuations. Simulations on the 24-site cluster were used to
determine the $\k$ dependence of $\Phid$.

\section{The Structure of the Pairing Interaction as Reflected in
\boldmath$\ld$ and \boldmath$\Phid$}
\label{sec:struct-pair-inter}

\begin{figure}[htpb]
	\begin{center}
          \includegraphics[width=3.5in]{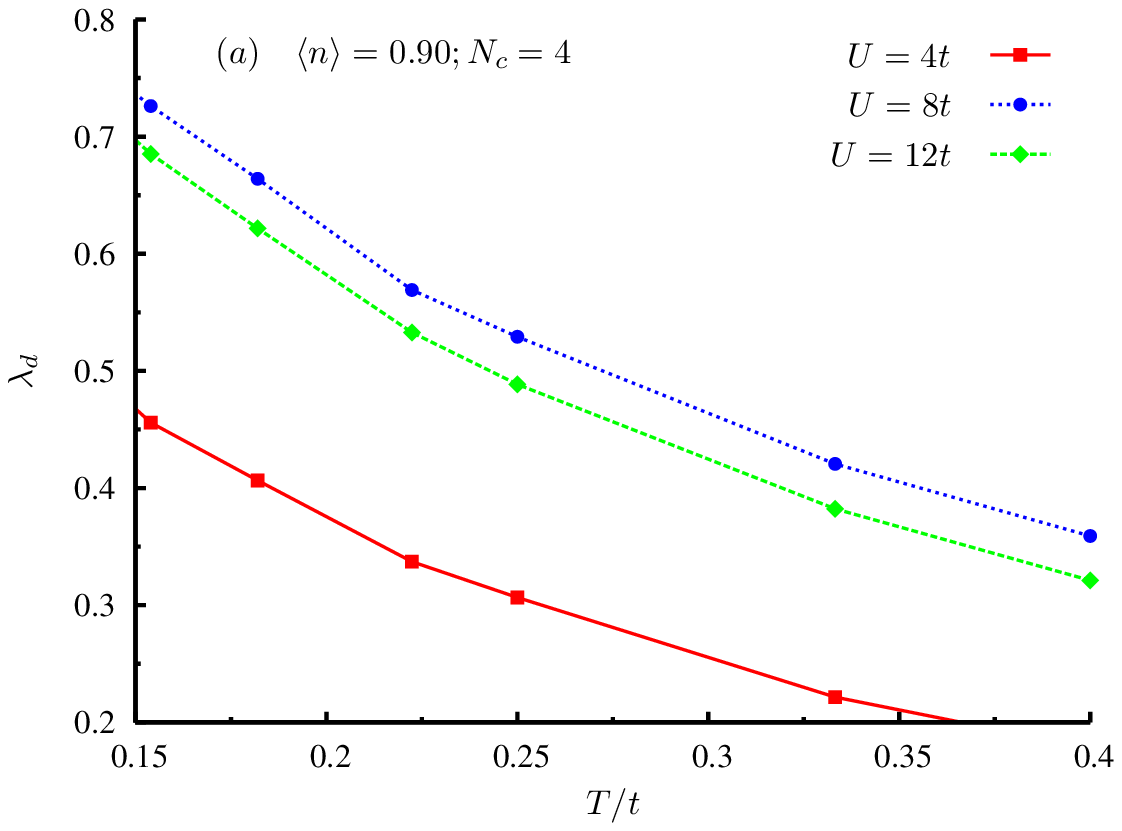}
          \includegraphics[width=3.5in]{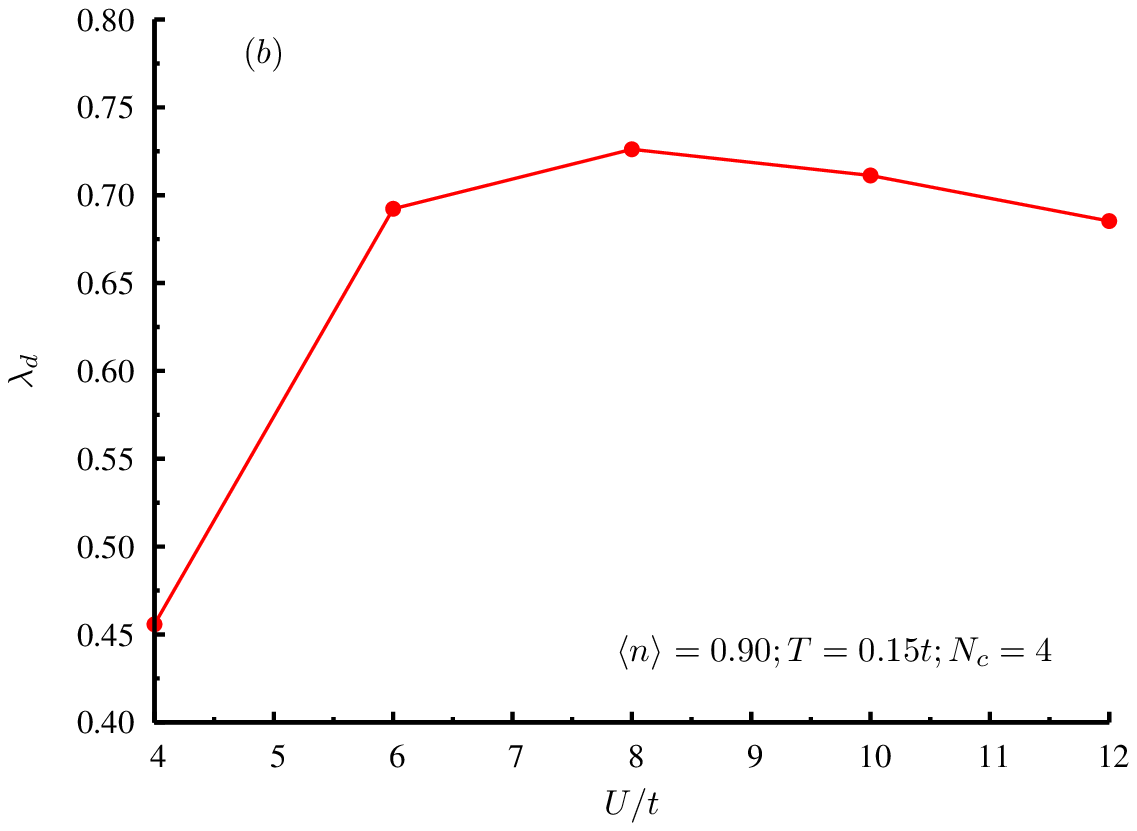}
	\end{center}
	\caption{(a) The $\dw$ eigenvalue $\ld$ versus $T/t$ for
          $U=4t$, $8t$ and $12t$ and $\n=0.90$. (b) The $\dw$
          eigenvalue $\ld$ versus $U/t$ for $T=0.15t$ and $\n=0.90$.}
	\label{fig:lambdad}
\end{figure}

The temperature dependence of the d-wave eigenvalues $\ld$ calculated using a
4-site cluster for a site
filling $\n=0.9$ with $U/t=4$, 8 and 12 are shown in
Fig.~\ref{fig:lambdad}a. Fig.~\ref{fig:lambdad}b shows $\lambda_d$
versus $U/t$ for $T=0.15t$.
Here one sees that it is favorable to have a Coulomb interaction
strength $U$ of order the bandwidth $8t$. This is consistent with the
notion that it is important to have strong short-range
antiferromagnetic correlations. However, because the exchange
interaction $J\sim 4t^2/U$ at strong coupling, the short-range antiferromagnetic
correlations decrease when $U$ becomes large compared to the bandwidth.

\begin{figure}[htpb]
	\begin{center}
          \includegraphics[width=3.5in]{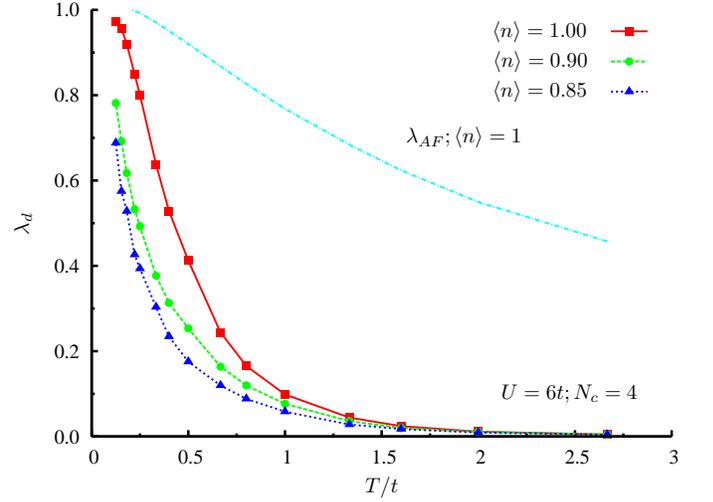}
	\end{center}
	\caption{The $\dw$ eigenvalue $\ld$ versus $T/t$ for various
          band fillings $\n$ for $U/t=6$. The dashed line represents
          the the leading eigenvalue $\lambda_{AF}$ in the
          $\Q=(\pi,\pi)$, $S=1$ particle-hole channel at half-filling.}
	\label{fig:lambdadn}
\end{figure}

The dependence of $\lambda_d(T)$ on the filling $\n$ is illustrated in
Fig.~\ref{fig:lambdadn} for $U/t=6$. What one sees is that $\lambda_d(T)$
increases as the system is doped towards half-filling. However, at
half-filling the dominant eigenvalue of the 4-point vertex occurs in the 
$\Q=(\pi,\pi)$ irreducible particle-hole $S=1$ channel as indicated
by the dashed line in Fig.~\ref{fig:lambdadn}, and the groundstate at
$T=0$ has long-range antiferromagnetic order.

\begin{figure}[htpb]
  \centering
  \includegraphics[width=3.5in]{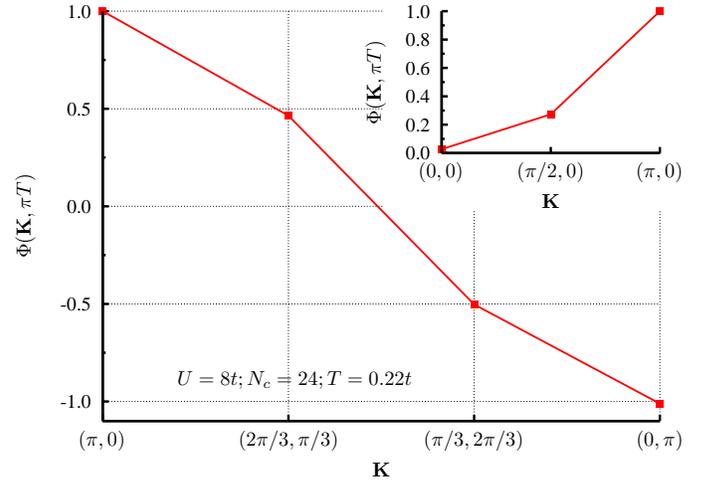}
  \caption{The $\dw$ eigenvector $\Phi_d(\K, \omega_n)$ at $\omega_n=\pi T$, 
normalized to its value at $\K = (\pi, 0)$,  versus
    $\K$ for $U/t=8$, band filling $\n=0.9$ and $T/t=0.22$. In the main figure, the
$\K$ points move along the dashed line shown in Fig.~3. The inset shows the
behavior of $\Phi_d$ when $\K$ varies along the $k_x$ axis.}
  \label{fig:PhipiTvsK}
\end{figure}

For $\n=0.9$ and $U/t=8$, the momentum dependence of the d-wave
eigenvector $\Phi_d(\K,\omega_n)$ for the 24-site cluster at $\omega_n=\pi T$
is shown in Fig.~\ref{fig:PhipiTvsK}. Here, for $T/t=0.22$, $\ld=0.42$ and the
values of $\K$ lay along the dashed line shown in Fig.~\ref{fig:DCABZ}. One
clearly sees the d-wave structure of $\Phi_d$.
The dependence of $\Phi_d(\K,\pi T)$ for $\K$ along the $K_x$ axis is
shown in the inset of Fig.~6. Here, one sees that $\Phi_d(\K,\pi T)$ falls off 
as $\K$ moves away from the Fermi surface towards the zone center. 

We have also calculated the projection of $\Phi_d(\K,\pi T)$ on the
first and second $\dw$ crystal harmonics
\begin{eqnarray}
\label{eq:1}
	d_i=\sum_{\K}g_i(\K)\Phi_d(\K,\pi T)
\end{eqnarray}
with $g_1(\K)=\cos K_x-\cos K_y$ and $g_2(\K)=\cos 2K_x-\cos 2K_y$.
In table~\ref{tab:harmonics}, we list the values of $d_2/d_1$ versus
$U$ at a filling $\n=0.9$. Here the sum in Eq.~(\ref{eq:1}) is over
the entire Brillouin zone and the temperature was adjusted so that the
d-wave eigenvalue $\lambda_d$ for each $U/t$ was the same
($\lambda_d\approx 0.4$). If the sum over $\K$ in Eq.~(\ref{eq:1}) is
restricted to values which lay along the dashed line in
Fig.~\ref{fig:DCABZ}, this ratio vanishes exactly in the 24-site cluster,
since $g_2(\K)=0$ on the momenta $\K$ along the dashed line. 
\begin{table}[htpb]
  \centering
  \begin{tabular}{c|ccc}
    $U/t$ & 4 & 6 & 8 \\\hline
    $d_2/d_1$ & 0.064 & 0.128 & 0.157
  \end{tabular}
  \caption{The ratio of the second to the first crystal d-wave
    harmonic projection of $\Phi_d(\K,\pi T)$ for $\n=0.9$ and
    $\lambda_d\approx 0.4$.}
  \label{tab:harmonics}
\end{table}

\begin{figure}[htpb]
  \centering
  \includegraphics[width=3.5in]{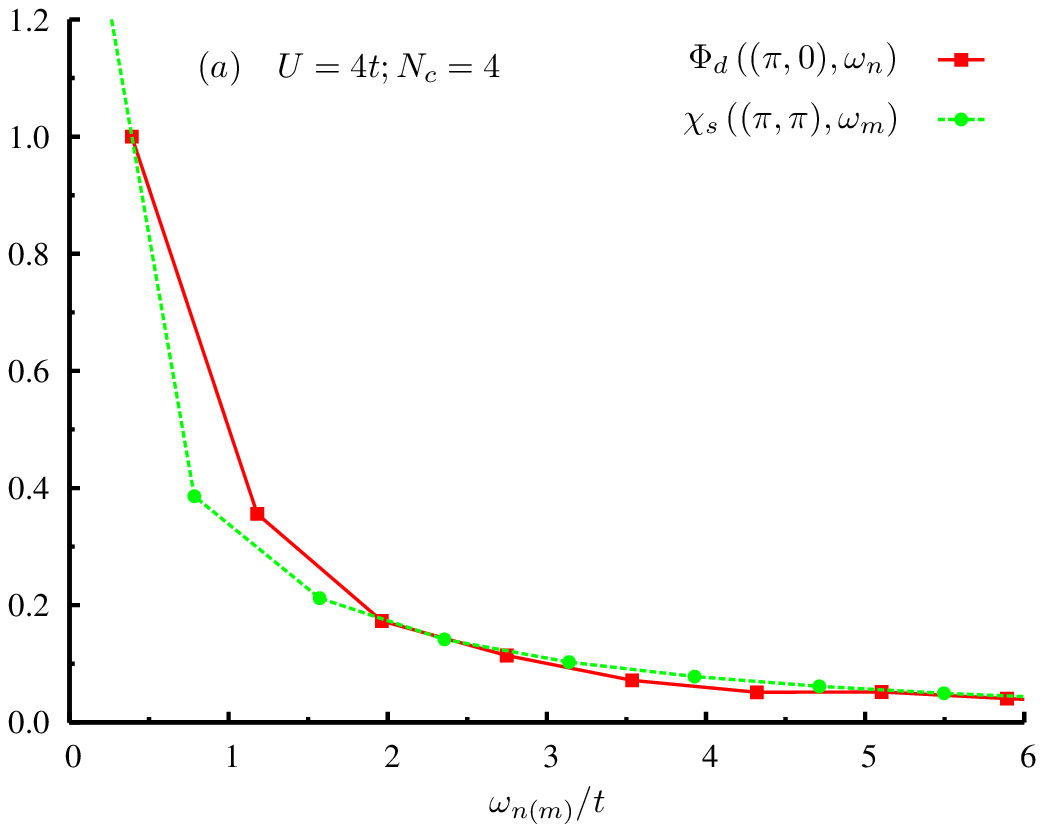}
  \includegraphics[width=3.5in]{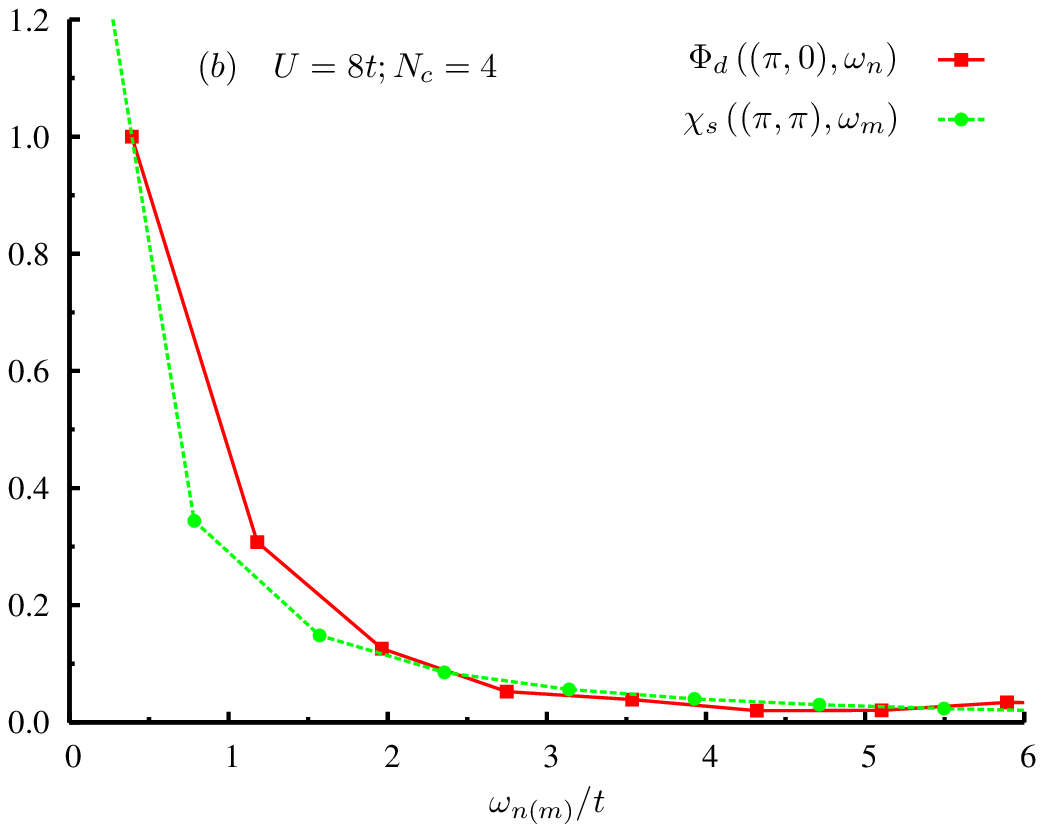}
  \includegraphics[width=3.5in]{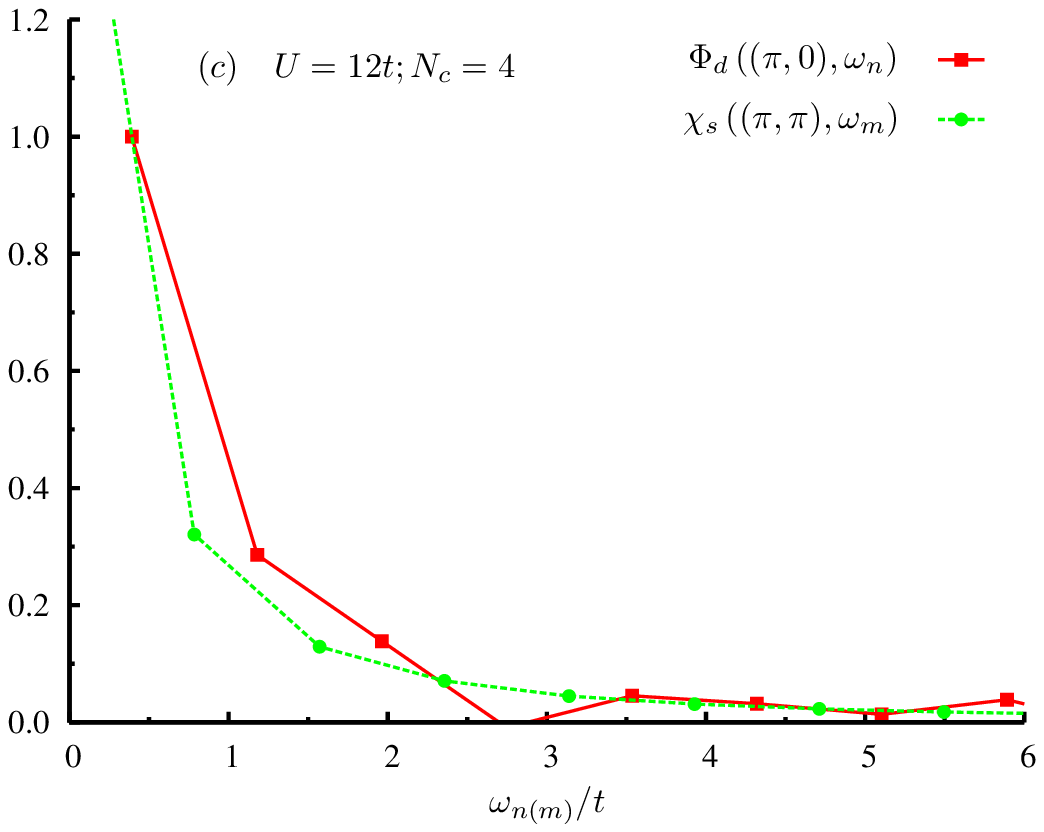}
  \caption{The Matsubara frequency dependence of
    $\Phi_d(\K,\omega_n)/\Phi_d(\K,\pi T)$ with $\K=(\pi,0)$ for (a)
    $U/t=4$, (b) $U/t=8$ and (c) $U/t=12$ calculated for $N_c$=4. Here
    $\omega_n=(2n+1)\pi T$ with $T/t=0.125$ and the band filling
    $\n=0.90$. Also shown is the frequency dependence of the
    normalized spin susceptibility
    $2\chi(\Q,\omega_m)/[\chi(\Q,0)+\chi(\Q,2\pi T)]$ for $\Q=(\pi,
    \pi)$.}
  \label{fig:PhiChi}
\end{figure}

The Matsubara frequency dependence of
$\Phi_d(\K,\omega_n)/\Phi_d(\K,\pi T)$ with $\K=(\pi,0)$ is shown in
Fig.~\ref{fig:PhiChi} for $\n=0.9$ and $U/t=4$, 8 and
12. Also shown in each case is the frequency dependence of
$\chi(\Q,\omega_n)$ with $\Q=(\pi,\pi)$. The decrease of the
characteristic exchange energy as $U/t$ increases is seen in 
Figs.~\ref{fig:PhiChi}a--c. It is clear from these
results that the dynamics of the pairing interaction $\Gpp$, reflected
in the $\omega_n$ dependence of $\Phid$, is associated with the
spin-fluctuation spectrum.

\section{A Separable Representation of the Pairing Interaction}
\label{sec:separ-repr-pair}

With the results obtained for the d-wave eigenvector $\Phi_d(K)$, one can
construct a simple separable representation of the pairing interaction
$\Gpp(K|K')$
\begin{equation}
\Gamma^{pp} (K| K^\prime)  \cong  -V_d\, \Phi_d(K)\, \Phi_d(K^\prime)\,.
\label{eq:ten}
\end{equation}
Using this separable form for $\Gamma^{pp} (K|K^\prime)$, one finds that
Eq.~(\ref{Eq:eigvalcg}) gives
\begin{equation}
  V_d \frac{T}{N_c}\ \sum_{K^\prime} \Phi_d^2(K^\prime)\,
  \bar{\chi}_0^{pp}(K^\prime) = \lambda_d\, .  
  \label{twelve}
\end{equation}
Then, using the dressed DCA Monte Carlo single particle Green's functions and
the DCA results for $\lambda_d$, we can determine the strength
$V_d$ of the separable interaction from Eq.~\eqref{twelve}. The
strength of $V_d$ depends upon both the site occupation $\langle
n\rangle$ and the temperature $T$. An alternative way of extracting a
simple approximation for the pairing interaction is to use the d-wave
projected irreducible vertex for $\omega_n=\omega_{n'}=\pi T$ 
\begin{eqnarray}
	V_d^{(\Gamma)}=\frac{1}{N_c}\sum_{\K,\K'}g(\K)\Gamma^{pp}(\K,\pi
	T|\K',\pi T)g(\K')
	\label{eq:Vdg}
\end{eqnarray}
with $g(\K)=(\cos K_x-\cos K_y)/2$.

\begin{figure}[htpb]
  \centering
  \includegraphics[width=3.5in]{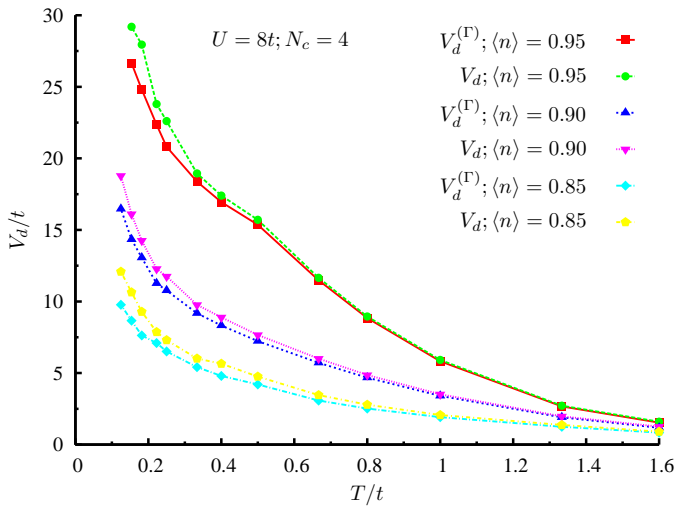}
  \caption{The pairing interaction $V_d$ versus temperature for
    different dopings for $U=8t$ calculated for $N_c=4$.}
  \label{fig:Vd}
\end{figure}

Fig.~\ref{fig:Vd} shows $V_d(T)$ and $V_d^{(\Gamma)}(T)$ for
$U=8t$ and various values of the site occupation $\n$. Both
approximations give very similar results. It is interesting to see that
$V_d$ becomes stronger as $\n$ approaches half-filling. This is
characteristic of the Mott-Hubbard system and has been previously
observed. A determinantal quantum Monte Carlo calculation of the d-wave
eigenvalue on an $8\times 8$ lattice at half-filling with $U=8t$ found
that $\lambda_d$ approached 1 as the temperature was lowered
\cite{BSW93}. As noted there, in this case the antiferromagnetic
eigenvalue remained dominant and on an infinite lattice the groundstate
would have long-range antiferromagnetic order at $T=0$. A related
behavior is seen for the two-leg ladder, where the pair binding energy
of a finite ladder is greatest for the first two holes that are added
\cite{SW00}. This latent pairing tendency of the half-filled Hubbard
model is also seen in the magnitude of the probability amplitude for
adding two near-neighbor holes in a d-wave state reported by Plekhanov
{\it et al.} \cite{PBS05}.

\begin{figure}[htpb]
  \centering
  \includegraphics[width=3.5in]{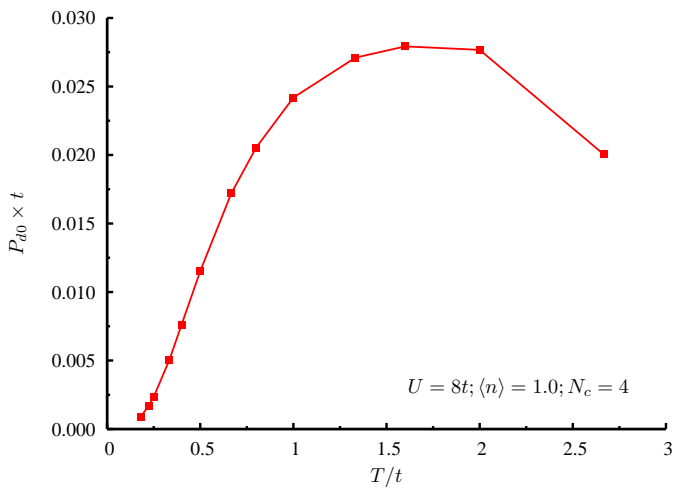}
  \caption{Plot of $P_{d0}(T)$ versus $T$ for $U=8t$ and $\n=1$
    showing the effect of the opening of the Mott-Hubbard gap}
  \label{fig:Pd0}
\end{figure}

Although $V_d$ increases as $\n$ goes to 1, the number of holes
that are available for pairing is suppressed as is $T_c$. A measure of this is
given by 
\begin{eqnarray}
\label{eq:Pd0}
P_{d0} (T)=\frac{T}{N_c}\ \sum_K \Phi_d(K)^2 \bar{\chi}_0^{pp}(K)\, 
\end{eqnarray}
which is plotted in Fig.~\ref{fig:Pd0} for $U=8t$ and $\n=1$. Here one
clearly sees that as the temperature is lowered and the Mott-Hubbard
gap opens, $P_{d0}(T)$ is suppressed.

\section{Conclusion}
\label{sec:conclusion}

The $\cos k_x-\cos k_y$ dependence of $\Phid$ reflects a pairing interaction
$\Gpp(\k|\k')$ which increases at large momentum transfer $\k-\k'$, implying a
spatially short-range interaction which is repulsive for pair formation on the
same site but attractive for singlet pair formation between near-neighbor
sites. The $\omega_n$ dependence of $\Phid$ tells us that the pairing
interaction is retarded on a time scale set by $J^{-1}$. The strength of the
interaction is largest when $U$ is of order the bandwidth and increases as the
system is doped towards half-filling. Of course, with $U=8t$, as $\n$ goes to
1, a Mott-Hubbard gap opens and there are no holes to pair.

\acknowledgments This research was enabled by computational resources of
the Center for Computational Sciences at Oak Ridge National Laboratory
and conducted at the Center for Nanophase Materials Sciences, which is
sponsored at Oak Ridge National Laboratory by the Division of Scientific
User Facilities, U.S. Department of Energy.  This research was supported
by NSF DMR-0312680.  DJS would like to thank the Stanford Applied
Physics Department for their hospitality.  DJS and MJ acknowledge the
Center for Nanophase Materials Science at Oak Ridge National Laboratory
for support.

\end{document}